\documentclass[12pt]{article} 
\usepackage[xdvi]{graphicx} 
\usepackage{amssymb}
 
\begin{document}   
\newcommand{\todo}[1]{{\em \small {#1}}\marginpar{$\Longleftarrow$}}   
\newcommand{\labell}[1]{\label{#1}\qquad_{#1}} 

\rightline{DCPT-04/03}   
\rightline{hep-th/0401205}   
\vskip 1cm


\begin{center} 
{\Large \bf Quotients of anti-de Sitter space}
\end{center} 
\vskip 1cm   
  
\renewcommand{\thefootnote}{\fnsymbol{footnote}}   
\centerline{\bf   
Owen Madden\footnote{O.F.Madden@durham.ac.uk} and Simon 
F. Ross\footnote{S.F.Ross@durham.ac.uk}}    
\vskip .5cm   
\centerline{ \it Centre for Particle Theory, Department of  
Mathematical Sciences}   
\centerline{\it University of Durham, South Road, Durham DH1 3LE, U.K.}   
  
\setcounter{footnote}{0}   
\renewcommand{\thefootnote}{\arabic{footnote}}


\begin{abstract}   
We study the quotients of $n+1$-dimensional anti-de Sitter space by
one-parameter subgroups of its isometry group $SO(2,n)$ for general
$n$. We classify the different quotients up to conjugation by
$O(2,n)$. We find that the majority of the classes exist for all
$n\geq 2$. There are two special classes which appear in higher
dimensions: one for $n \geq 3$ and one for $n \geq 4$. The description
of the quotient in the majority of cases is thus a simple
generalisation of the AdS$_3$ quotients. 
\end{abstract}    
  
The study of the propagation of strings on more general curved
backgrounds is important both because it allows us to confront some of
the important problems arising in any theory of quantum gravity (such
as the problem of time), and because describing strings on
time-dependent backgrounds is essential to address the
phenomenological application of string theory to cosmology. A new
class of simple supersymmetric backgrounds referred to as null branes
was recently constructed~\cite{fofs}, by considering a novel class of
Kaluza-Klein reductions of flat space. These do not have a timelike
Killing field, so they provide interesting examples for studying
string theory on more general backgrounds; in addition, a subclass of
`parabolic orbifolds' have initial singularities. String theory on
these backgrounds has been intensively studied, to expand our
understanding of string theory in non-static backgrounds and to
attempt to gain insight into the resolution of such spacetime
singularities in string theory~\cite{lms1,lms2,fabmcg}. Unfortunately,
unlike in more familiar spacelike orbifolds, it turns out that the
singular geometries suffer from an instability, so the resolution of
the singularities is not accessible in perturbation theory
\cite{lawrence,lms2,fabmcg,horpol}.

It is natural for many reasons to wish to extend these investigations
to consider strings on orbifolds of Anti-de Sitter space (AdS). First,
AdS is also a maximally symmetric space, so it has a large isometry
group which can lead to interesting examples of quotients. Secondly,
the AdS/CFT correspondence~\cite{mald,magoo} provides a
non-perturbative definition of string theory, which may enable us to
obtain more insight into issues such as singularity resolution in an
AdS context. Finally, it is well-known that a black hole geometry can
be constructed from a quotient of AdS$_3$~\cite{ban1,ban2}. These
constructions therefore also offer an opportunity to explore
backgrounds with non-trivial causal structure.

Such an extension was initiated in~\cite{simon}, where an AdS version
of the isometry involved in the null brane quotient was constructed.
Our aim in the present paper is to make a more systematic
investigation of this question, classifying all the physically
distinct quotients of AdS$_{n+1}$ by one-parameter subgroups of its
isometry group. The classification of quotients of AdS$_3$ was
thoroughly explored in~\cite{btz2}. This was extended to AdS$_4$
in~\cite{hp}. Our aim is to extend this to general dimensions, and in
particular to address the case of AdS$_5$, of great interest for
string theory. This question has also been explored independently by
Figueroa-O'Farrill and Simon~\cite{fofs2}, who also investigate
quotients with a non-trivial action on the sphere factor in AdS$_p
\times S^q$ backgrounds in string theory and investigate the
supersymmetry preserved under their quotients.

We will show that the classification of physically distinct
one-parameter subgroups of $SO(2,n)$ extends very naturally from the
case $n=2$ to higher $n$. The subgroups considered in~\cite{btz2} all
have higher-dimensional generalisations, whose analysis is directly
related to the analysis in the case of AdS$_3$. There are only two
further physically distinct possibilities, one of which appears for
all $n \geq 3$, and the other of which appears for all $n \geq 4$. The
prototype example of the former was discussed in~\cite{hp}, and the
latter contains the null brane-like quotient discussed
in~\cite{simon}.

The purpose of this paper is to describe the basic steps in the
classification of the quotients and the construction of normal forms
for the Killing vectors in some detail.  We will also briefly explore
how the coordinate systems can be adapted to directly relate
higher-dimensional quotients to lower-dimensional ones, but we
postpone detailed exploration of the physics of these quotients to a
companion paper with Figueroa-O'Farrill and Simon~\cite{collab}.

We wish to classify quotients of AdS$_{n+1}$ by one-parameter
subgroups of $SO(2,n)$.\footnote{We will generally have in mind the
quotient by a discrete subgroup, to construct another
$n+1$-dimensional spacetime; the prototypical example is the BTZ black
hole~\cite{btz1,btz2}.  It is also interesting to consider the
Kaluza-Klein reduction along such a direction to construct an
$n$-dimensional spacetime. For the purposes of classification, we can
treat these two kinds of quotients together.} A one-parameter subgroup
is determined by a Killing vector $\xi^\mu$ in the Lie algebra
$so(2,n)$; such a Killing vector can be written in terms of a basis
$J_{ab}^\mu$ of $so(2,n)$ as $\xi = \omega^{ab} J_{ab}$, where
$\omega^{ab} = -\omega^{ba}$. If we describe AdS$_{n+1}$ in terms of
embedding coordinates $U,V,X_i$ ($i=2,\ldots, n+1$) such that $-U^2 -
V^2 + X_i^2 = -1$, then the $J_{ab}$ are
\begin{equation}
J_{01} = V \partial_U - U \partial_V, \quad J_{0i} = U \partial_i +
X_i \partial_U, \quad J_{1i} = V \partial_i + X_i \partial_V, \quad
J_{ij} = X_i \partial_j - X_j \partial_i. 
\end{equation}

The classification of physically different $\xi^\mu$ is therefore
equivalent to classifying antisymmetric matrices $\omega^{ab}$ up to
conjugation by elements of $SO(2,n)$, that is, $\omega' \sim \omega$
iff $\omega'^a_{\ b} = (T^{-1})^a_{\ c} \omega^c_{\ d} T^d_{\ b}$ for
some $T^a_{\ c} \in SO(2,n)$. As explained in~\cite{btz2,hp}, if we
slightly extend the equivalence relation, so that $\omega' \sim
\omega$ for $T^a_{\ c} \in O(2,n)$, then the problem is equivalent to
the familiar problem of classifying the matrices up to similarity. The
distinct matrices are then classified by their eigenvalues and the
dimensions of the irreducible invariant subspaces associated with
them. This extension of the equivalence relation implies that we will
not distinguish between Killing vectors which differ by a sign
reversal of some of the embedding coordinates. 

Since the classification reduces to the study of the eigenvalues and
eigenspaces of the matrix $\omega^a_{\ b}$, we can `build up' the
general matrix from the different eigenspaces. We will therefore first
consider the different possibilities for invariant subspaces
consistent with the signature of spacetime, and then use these
possible invariant subspaces as building blocks to construct all the
possible inequivalent matrices $\omega_{ab}$, and hence classify the
different quotients. In the following we shall say that the matrix
$\omega_{ab}$ is of type $k$ if its highest dimensional irreducible
invariant subspace is of dimension $k$.

The calculations are simplified by observing that as a consequence of
the fact that $\omega_{ab}$ is real and antisymmetric, its eigenvalues
come in groups: if $\lambda$ is an eigenvalue of $\omega^a_{\ b}$ then
$-\lambda$ is an eigenvalue of $\omega^a_{\ b}$, and similarly if
$\lambda$ is an eigenvalue then so is $\lambda^{*}$. Another useful
fact is that if $v^{a}$ and $u^{a}$ are eigenvectors of
$\omega^{a}_{\ b}$ with respective eigenvalues $\lambda$ and $\mu$, so
that 
\begin{equation}
\omega^{a}_{\ b}v^{b}=\lambda v^{a}, \qquad \omega^{a}_{\ b}u^{b}=\mu
u^{a},
\end{equation}
then $v^a u_a=0$ unless $\lambda +\mu=0$. Note that $v^a$ etc. are
vectors in $\mathbb{R}^{2,n}$; the indices on $\omega_{ab}$, $v^a$ etc
are raised and lowered with the metric $\eta_{ab}$ on
$\mathbb{R}^{2,n}$. Thus, we see that $\mathbb{R}^{2,n}$ decomposes
into a product of orthogonal eigenspaces, but each such subspace is
associated not with a single eigenvalue $\lambda$ but with the pair of
eigenvalues $\lambda, -\lambda$. We will now study the properties of
these orthogonal eigenspaces.

Let us first discuss the cases with non-degenerate eigenvalues. The
simplest case is when the eigenvalue is zero; then there is a single
eigenvector $v^a$, which is orthogonal to all other eigenvectors, and
by the non-degeneracy of the metric must then have $v^a v_a \neq
0$. We can rescale $v^a$ to set $v^a v_a = 1$, which we will refer to
as $\lambda_0(+)$, or $v^a v_a = -1$, which we will refer to as
$\lambda_0(-)$. These cases correspond physically to a direction in
$\mathbb{R}^{2,n}$ which is not affected by the identification.

The next possibility is a pair of real eigenvalues, $a,-a$, $a \geq
0$. Then we have
\begin{equation}
\omega_{ab} l^b = a l_a, \quad \omega_{ab} m^b = -a m_a. 
\end{equation}
The only non-zero inner product is $l_a m^a = 1$. To construct an
orthonormal basis, we take
\begin{equation}
v_1 = {1 \over \sqrt{2}} (l+m), \quad v_2 = {1 \over \sqrt{2}} (l-m).
\end{equation}
We then have $v_1 \cdot v_1= 1$, $v_2 \cdot v_2 = -1$, so this
subspace has signature $(-+)$. We denote this by $\lambda_r$; it
corresponds physically to a boost in some $\mathbb{R}^{1,1}$ subspace
of $\mathbb{R}^{2,n}$. 

If we have a pair of imaginary eigenvalues, 
\begin{equation}
\omega_{ab} k^b = ib k_a, \quad \omega_{ab} k^{*b} = -ib k^*_a, 
\end{equation}
$b \geq 0$, the only non-zero inner product is $k_a k^{*a} = 1$. Now
we need to construct the orthonormal basis in a slightly different
way, because we need to respect the fact that the action of
$\omega_{ab}$ on $\mathbb{R}^{2,n}$ is real-valued. We can set
\begin{equation}
v_1 = {1 \over \sqrt{2}} (k+k^*), \quad v_2 = {i \over \sqrt{2}} (k-k^*).
\end{equation}
We then have $\omega^a_{\ b} v_1^b = b v_2^a$, $\omega^a_{\ b} v_2^b = -b
v_1^a$. We have $v_1 \cdot v_1= 1$, $v_2 \cdot v_2 = 1$, so this
subspace has signature $(++)$, which we denote by $\lambda_i (++)$. On
the other hand, we could have chosen 
\begin{equation}
v_1 = {i \over \sqrt{2}} (k+k^*), \quad v_2 = {1 \over \sqrt{2}} (k-k^*).
\end{equation}
This also gives a real action, but now $v_1 \cdot v_1= -1$, $v_2 \cdot
v_2 = -1$, so this subspace has signature $(--)$, which we denote by
$\lambda_i (--)$. These two cases correspond physically to rotations
in $\mathbb{R}^2$ subspaces of $\mathbb{R}^{2,n}$. 

The final possibility is a complex eigenvalue, which gives us the four
eigenvalues $\lambda, -\lambda, \lambda^*, -\lambda^*$ (so we can take
$\lambda = a+ib$ for $a,b \geq 0$). We have
\begin{equation}
\omega_{ab} l^b = \lambda l_a, \quad \omega_{ab} m^{b} = -\lambda m_a, 
\end{equation}
\begin{equation}
\omega_{ab} l^{*b} = \lambda^* l^*_a, \quad \omega_{ab} m^{*b} =
-\lambda^* m^*_a.   
\end{equation}
The non-vanishing inner products are $l \cdot m = 1$ and $l^* \cdot
m^* = 1$, so $l,m$ and $l^*,m^*$ span two orthogonal two-dimensional
spaces; however, we need to mix them to obtain a real basis. If we
define
\begin{equation}
v_1 = {1 \over 2} [(l+l^*)+(m+m^*)], \quad v_2 = {1 \over 2}
[(l+l^*)-(m+m^*)], 
\end{equation}
\begin{equation}
v_3 = {i \over 2} [(l-l^*)+(m-m^*)], \quad v_4 = {i \over 2}
[(l-l^*)-(m-m^*)],  
\end{equation}
Then we will see that $\omega_{ab}$ acts on the $v_i$ with real
coefficients, and they span a space of signature $(--++)$, which we
denote $\lambda_c$. 

Now we turn to the higher-dimensional invariant subspaces. If we have
a $k$-dimensional subspace associated to the eigenvalue zero, then we
can pick a basis of vectors $m_i$, $i=1,\ldots,k$ such that 
\begin{equation} \label{omega0}
\omega_{ab} m_1^b = 0, \quad \omega_{ab} m_i^b = m_{(i-1)a} \mbox{ for
} i \neq 1.  
\end{equation}
We can then observe that $m_1^a m_{(i-1)a} = m_1^a \omega_{ab} m_i^b =
0$ for $i = 1,\ldots,k$. We then need $m_1^a m_{ka} \neq 0$ for
consistency with the non-degenerate metric. We can also use
(\ref{omega0}) to show
\begin{equation} \label{recurs0}
m_{ia} m_j^a = m_{ia} \omega^{ab} m_{(j+1)b} = - m_{(i-1)a} m_{(j+1)}^a, 
\end{equation}
and
\begin{equation}
m_{ia} m_{(i-1)}^a = m_{ia} \omega^{ab} m_{ib} = 0 
\end{equation}
by antisymmetry of $\omega_{ab}$. Now imagine $k$ is even. Then these
two relations taken together imply that 
\begin{equation}
m_{ka} m_1^a = m_{(k/2)a} m_{(k/2+1)}^a = 0, 
\end{equation}
in contradiction with the non-degeneracy of the metric. Hence there
cannot be $k$-dimensional invariant subspaces associated with a zero
eigenvalue for $k$ even. For $k$ odd, (\ref{recurs0}) implies
\begin{equation}
m_{ia} m_j^a = (-1)^{i+1} m_{1a} m_k^a
\end{equation}
for $i+j = k+1$. We can also set all other inner products to zero by a
suitable redefinition of the basis $m_i^a$. We can then define an
orthonormal basis by
\begin{equation}
v_{2i-1} = {1 \over \sqrt{2}} (m_i + m_{k+1-i}), \quad v_{2i} = {1
\over \sqrt{2}} (m_i - m_{k+1-i}) \mbox{ for } i = 1, \ldots, {k-1
\over 2}, 
\end{equation}
and $v_k = m_{k+1/2}$. We then have $v_{2i-1} \cdot v_{2i-1} = -
v_{2i} \cdot v_{2i}$, and we can choose $v_k \cdot v_k$ to be $\pm 1$,
so the subspace spanned by these vectors has either $(k-1)/2$ negative
signature directions and $(k+1)/2$ positive signature ones, or
$(k+1)/2$ negative signature directions and $(k-1)/2$ positive
signature ones. The only possibilities which are consistent with
embedding as a subspace in $\mathbb{R}^{2,n}$ are $\lambda_0^{III} (-++)$
and $\lambda_0^{III} (--+)$, and $\lambda_0^V$ with signature
$(--+++)$. $\lambda_0^{III}$ corresponds to a null rotation in an
$\mathbb{R}^{1,2}$ subspace of $\mathbb{R}^{2,n}$. 

If we have a $k$-dimensional invariant subspace with a real
eigenvalue, we must have a pair of them; we can define a basis such
that the action of $\omega_{ab}$ is
\begin{equation}
\omega_{ab} l_1^b = a l_{1a}, \quad \omega_{ab} l_i^b = a l_{ia} +
l_{(i-1) a}\mbox{ for
} i =2, \ldots, k  ,
\end{equation}
and
\begin{equation}
\omega_{ab} m_1^b = -a m_{1a}, \quad \omega_{ab} m_i^b = -a m_{ia} +
m_{(i-1) a}\mbox{ for
} i =2, \ldots, k .
\end{equation}
By repeatedly using these relations, we can show that $l_i \cdot l_j =
0$ and $m_i \cdot m_j = 0$ for all $i,j$. We can also show $m_1 \cdot
l_i = 0$ for $i \neq k$; we then need $m_1 \cdot l_k \neq 0$ for
non-degeneracy. As in the case of a zero eigenvalue, we learn that
\begin{equation}
m_i \cdot l_j = (-1)^{i+1} m_1 \cdot l_k,
\end{equation}
for $i+j = k+1$, and we can set all other inner products to zero by a
suitable redefinition of the basis. An orthonormal basis is then
formed by taking
\begin{equation}
v_{2i-1} = {1 \over \sqrt{2}} (l_i + m_{k+1-i}), \quad v_{2i} = {1
\over \sqrt{2}} (l_i - m_{k+1-i}) \mbox{ for } i = 1, \ldots, k. 
\end{equation}
We then have $v_{2i-1} \cdot v_{2i-1} = - v_{2i} \cdot v_{2i}$, so the
subspace spanned by these vectors has an equal number of negative and
positive signature directions. The only possibility consistent with
being a subspace of $\mathbb{R}^{2,n}$ is $\lambda_r^{II}$, which has
signature $(--++)$. 

If we have a $k$-dimensional invariant subspace with an imaginary
eigenvalue, we must again have a pair of them; we can define a basis such
that the action of $\omega_{ab}$ is
\begin{equation}
\omega_{ab} k_1^b = ib k_{1a}, \quad \omega_{ab} k_i^b = ib k_{ia} +
k_{(i-1) a}\mbox{ for } i =2, \ldots, k ,
\end{equation}
and
\begin{equation}
\omega_{ab} k_1^{*b} = -ib k^*_{1a}, \quad \omega_{ab} k_i^{*b} = -ib
k^*_{ia} + k^*_{(i-1) a}\mbox{ for } i =2, \ldots, k .
\end{equation}
By repeatedly using these relations, we can show that $k_i \cdot k_j =
0$ and $k^*_i \cdot k^*_j = 0$ for all $i,j$. We can also show $k_1 \cdot
k^*_i = 0$ for $i \neq k$; we then need $k_1 \cdot k^*_k \neq 0$ for
non-degeneracy. As in the case of a zero eigenvalue, we learn that
\begin{equation}
k_i \cdot k^*_j = (-1)^{i+1} k_1 \cdot k^*_k
\end{equation}
for $i+j = k+1$, and we can set all other inner products to zero by a
suitable redefinition of the basis. The action of $\omega$ becomes
real if we define new vectors $w_i = {1 \over \sqrt{2}} (k_i + k_i^*)$
and $x_i = {i \over \sqrt{2}} (k_i - k_i^*)$. There is then a
technical difference between even and odd dimensions: in even
dimensions, the non-zero inner products are $w_i \cdot x_j$ for $i+j =
k+1$, and an orthonormal basis is formed by
taking 
\begin{equation}
v_{2i-1} = {1 \over \sqrt{2}} (w_i + x_{k+1-i}), \quad v_{2i} = {1
\over \sqrt{2}} (w_i - x_{k+1-i}) \mbox{ for } i = 1, \ldots, k, 
\end{equation}
We then have $v_{2i-1} \cdot v_{2i-1} = - v_{2i} \cdot v_{2i}$. Thus,
in even dimensions, we have a subspace with an equal number of
positive and negative directions, and the only possibility in
$\mathbb{R}^{2,n}$ is $\lambda_i^{II}$, which has signature $(--++)$. In
odd dimensions, the non-zero inner products are $w_i \cdot w_j = x_i
\cdot x_j$ for $i+j = k+1$, and an orthonormal basis is formed by
\begin{equation}
v_{2i-1} = {1 \over \sqrt{2}} (w_i + w_{k+1-i}), \quad v_{2i} = {1
\over \sqrt{2}} (w_i - w_{k+1-i}) \mbox{ for } i = 1, \ldots, {k-1
\over 2}, 
\end{equation}
\begin{equation}
v_{k} = w_{k+1 \over 2}, \quad v_{k+1} = x_{k+1 \over 2}
\end{equation}
\begin{equation}
v_{2i-1} = {1 \over \sqrt{2}} (x_i + x_{k+1-i}), \quad v_{2i} = {1
\over \sqrt{2}} (x_i - x_{k+1-i}) \mbox{ for } i = {k +3\over 2}, \ldots, k.
\end{equation}
We then have $v_{2i-1} \cdot v_{2i-1} = - v_{2i} \cdot v_{2i}$ except
for $i ={k+1 \over 2}$; $v_k \cdot v_k = v_{k+1} \cdot v_{k+1}$.  The
subspace thus either has $k-1$ positive and $k+1$ negative directions
or vice-versa. The only possibility in $\mathbb{R}^{2,n}$ is
$\lambda_i^{III}$, which has signature $(--++++)$. In the special case
$b=0$, which will be important later, $\lambda_i^{III}$ reduces to a
pair of $\lambda_0^{III} (-++)$---that is, to a pair of null rotations
in independent subspaces. 
Finally, we could consider invariant subspaces of dimension $k$
associated with complex eigenvalues. We will not give the details
here, as it does not lead to any cases that fit inside
$\mathbb{R}^{2,d}$. The subspace associated with the set of four
complex eigenvalues always has at least $2k$ negative directions. 

In summary, the possible invariant subspaces and their signatures that
can occur in our $\omega_{ab}$ are $\lambda_0(+)$, $\lambda_0(-)$,
$\lambda_r (-+)$, $\lambda_i (++)$, $\lambda_i (--)$, $\lambda_c
(--++)$, $\lambda_0^{III} (-++)$, $\lambda_0^{III} (--+)$,
$\lambda_0^V (--+++)$, $\lambda_r^{II} (--++)$, $\lambda_i^{II}
(--++)$, and $\lambda_i^{III} (--++++)$.\footnote{Naturally, the same
classification can be applied for the Lorentz group $SO(1,n)$ in
$\mathbb{R}^{1,n}$; in that case, the only possible subspaces are
$\lambda_0(+)$, $\lambda_0(-)$, $\lambda_r (-+)$, $\lambda_i (++)$,
and $\lambda_0^{III} (-++)$, corresponding to trivial directions,
boosts, rotations and null rotations respectively.} Now let us
consider how we can assemble these to form an $n+2$ dimensional matrix
$\omega_{ab}$.  For $n$ even (which includes the case $n=4$ which we
are particularly interested in), the possibilities are
\begin{itemize} 
\item
Type $I$
\eqnarray
&\mathbb{C}& \qquad \lambda_{c}~(--++)\qquad+\frac{n-2}{2}\lambda_{i}
~(+_{1}+_{2}\cdots+_{n-2}), \nonumber\\
&\mathbb{R}& \qquad 2\lambda_{r}~(--++)\qquad+\frac{n-2}{2}\lambda_{i}
~(+_{1}+_{2}\cdots+_{n-2}), \nonumber\\
&\mathbb{I}& \qquad \frac{n+2}{2}\lambda_{i}
~(--+_{1}+_{2}\cdots+_{n}). \nonumber
 \endeqnarray
Where the coefficient in front of a $\lambda$ corresponds to the number of 
times that type of eigenvalue appears.

\item
 Type $II$
\eqnarray
&\mathbb{R}& \qquad \lambda^{II}_{r}~(--++)\qquad+\frac{n-2}{2}\lambda_{i}
~(+_{1}+_{2}\cdots+_{n-2}), \nonumber\\
&\mathbb{I}& \qquad \lambda^{II}_{i}~(--++)\qquad+ \frac{n-2}{2}\lambda_{i}
~(+_{1}+_{2}\cdots+_{n-2}) \nonumber.
\endeqnarray

\item
Type $III$
\eqnarray
&\mathbb{I}& \qquad\lambda^{III}_{i}~(--++++)\qquad+ \frac{n-4}{2}\lambda_{i}
~(+_{1}+_{2}\cdots+_{n-4}), \nonumber\\
&0&(a)\qquad\lambda^{III}_{0}~(-++)\qquad+\lambda_{0}~(+)\qquad+\lambda_{r}~(-+)
\qquad+ \frac{n-4}{2}\lambda_{i}
~(+_{1}+_{2}\cdots+_{n-4}), \nonumber\\
&0&(b)\qquad\lambda^{III}_{0}~(--+)\qquad+\lambda_{0}~(+)\qquad+ \frac{n-2}{2}\lambda_{i}
~(+_{1}+_{2}\cdots+_{n-2}), \nonumber\\
&0&(c)\qquad\lambda^{III}_{0}~(-++)\qquad+\lambda_{0}~(-)\qquad+ \frac{n-2}{2}\lambda_{i}
~(+_{1}+_{2}\cdots+_{n-2}) \nonumber.
\endeqnarray
\item
Type $V$
\eqnarray
\lambda^{V}_{0}~(--+++)\qquad+\lambda_{0}~(+)\qquad+ \frac{n-4}{2}\lambda_{i}
~(+_{1}+_{2}\cdots+_{n-4}) \nonumber.
\endeqnarray
\end{itemize}

To discuss the physics of these different cases, we need a convenient
representative of each case. It is easy to construct suitable
representatives; in most cases, this is a minor generalisation of the
analysis of~\cite{btz2,hp}, so we will just quote the result by giving
the relevant Killing vectors. For $I_{\mathbb{C}}$ this is
\begin{equation}
\xi=b_{1}(J_{01}+J_{23})-a(J_{03}+J_{12})+b_{2}J_{45}+b_{3}J_{67}+\cdots
+b_{\frac{n}{2}}J_{nn+1},
\end{equation}
with $a, b_i \geq 0$.\footnote{Recall that we have identified Killing
vectors differing by conjugation by $O(2,n)$; if we only identified
under conjugation by $SO(2,n)$, we should take $b_i, i \geq 2$ to run
over the reals, and
$\xi=b_{1}(-J_{01}+J_{23})-a(-J_{03}+J_{12})+b_{2}J_{45}+b_{3}J_{67}+\cdots
+b_{\frac{n}{2}}J_{nn+1}$ and
$\xi=b_{1}(-J_{01}+J_{23})-a(J_{03}-J_{12})+b_{2}J_{45}+b_{3}J_{67}+\cdots
+b_{\frac{n}{2}}J_{nn+1}$ for $a, b_1 \geq 0$ would also count as
distinct cases. Similar remarks apply in the other cases to follow.}
The norm of this Killing vector is
\begin{eqnarray}
\xi_{\mu}\xi^{\mu}&=&(a^{2}-b_{1}^{2})
(X_{n+1}^{2}+X_{n}^{2}+\cdots+X_{4}^{2}+1)-4ab_{1}(V X_{3}- U X_{2})
\nonumber \\
&&+b_{2}^{2}(X_{4}^{2}+X_{5}^{2})+b_{3}^{2}(X_{6}^{2}+X_{7}^{2})+\cdots+
b_{\frac{n}{2}}^{2}(X_{n}^{2}+X_{n+1}^{2}).
\end{eqnarray}
Thus, this Killing vector can be everywhere spacelike for $b_1=0$. For
type $I_{\mathbb{R}}$ we have
\begin{equation}
\xi=a_{1}J_{03}+a_{2}J_{12}+b_{1}J_{45}+\cdots +b_{\frac{n-2}{2}}J_{nn+1},
\end{equation}
with norm
\begin{equation}
\xi_{\mu}\xi^{\mu}=a_{1}^{2}(U^{2}-X_{3}^{2})+
a_{2}^{2}(V^{2}-X_{2}^{2})+b_{1}^{2}(X_{4}^{2}+X_{5}^{2})
+\cdots+
b_{\frac{n-2}{2}}^{2}(X_{n}^{2}+X_{n+1}^{2}).
\end{equation}
This is everywhere spacelike for $a_1 = a_2$ (using $\eta^{ab} X_a X_b
= -1$), which is equivalent to type $I_{\mathbb{C}}$ with $b_1=0$. For
type $I_{\mathbb{I}}$ we have
\begin{equation}
\xi=b_{1}J_{01}+b_{2}J_{23}+b_{3}J_{45}+\cdots
+b_{\frac{n+2}{2}}J_{nn+1},
\end{equation}
with norm
\begin{equation}
\xi_{\mu}\xi^{\mu}=b_{1}^{2}(-1 - X_2^2 - \cdots -
X_{n+1}^2)+b_{2}^{2}(X_{2}^{2}+X_{3}^{2})+
b_{3}^{2}(X_{4}^{2}+X_{5}^{2})+\cdots+
b_{\frac{n+2}{2}}^{2}(X_{n}^{2}+X_{n+1}^{2}).
\end{equation}
For $b_1=0$, this is spacelike away from the axis $X_i = 0, i \geq 2$,
where the Killing vector degenerates, so this axis is a line of fixed
points. For type $II_{\mathbb{R}}$ we have
\begin{equation}
\xi=a(J_{03}+J_{12})+J_{01}-J_{02}-J_{13}+J_{23}+b_{1}J_{45}+\cdots
+b_{\frac{n-2}{2}}J_{nn+1},
\end{equation}
with norm
\begin{eqnarray}
\xi_{\mu}\xi^{\mu}&=&a^{2}(U^{2}+V^{2}-X_{2}^{2}-X_{3}^{2})+
4a(U-X_{3})(X_{2}+V)\nonumber \\
&&+b_{1}^{2}(X_{4}^{2}+X_{5}^{2})+\cdots+ 
b_{\frac{n-2}{2}}^{2}(X_{n}^{2}+X_{n+1}^{2}). 
\end{eqnarray}
For $a=0$, this is spacelike except on the subspace $X_i = 0, i \geq
4$, where the Killing vector is null. For type $II_{\mathbb{I}}$, we
have
\begin{equation}
\xi=(b_{1}-1)J_{01}+(b_{1}+1)J_{23}+J_{02}-J_{13}
+b_{2}J_{45}+\cdots+b_{\frac{n}{2}}J_{nn+1}, 
\end{equation}
with norm 
\begin{eqnarray}
\xi_{\mu}\xi^{\mu}&=&
b_{1}^{2}(-1 - X_4^2 - \cdots - X_{n+1}^2) \nonumber \\
&&+2b_{1}(U+X_3)^2 + 2b_1 (V+X_{2})^{2} \nonumber \\
&&+b_{2}^{2}(X_{4}^{2}+X_{5}^{2}) +\cdots+
b_{\frac{n-2}{2}}^{2}(X_{n}^{2}+X_{n+1}^{2}). 
\end{eqnarray}
For $b_1 = 0$, this is the same as type $II_{\mathbb{R}}$ with $a=0$
(as one would expect). For type $III_{\mathbb{I}}$ we have
\begin{equation}
\xi = b(J_{01}+ J_{23} + J_{45}) - J_{04} + J_{34} + J_{15} - J_{25}
+b_{2}J_{67}+\cdots+b_{\frac{n-2}{2}}J_{nn+1}, 
\end{equation}
with norm
\begin{eqnarray}
\xi_{\mu}\xi^{\mu} &=& -b^2 -4b(X_5(X_2-V)+X_4(X_3-U)) + (U-X_3)^2
+ (V-X_2)^2  \nonumber \\
&&+ b_{2}^{2}(X_{6}^{2}+X_{7}^{2})+\cdots+ b_{\frac{n-2}{2}}^{2}(X_{n}^{2}+X_{n+1}^{2}).
\end{eqnarray}
This is everywhere spacelike if $b=0$. For type $III_{0(a)}$ we have
\begin{equation}
\xi=-aJ_{15}-J_{03}+J_{23}+b_{1}J_{67}+\cdots+b_{\frac{n-4}{2}}J_{nn+1},
\end{equation}
with norm
\begin{equation}
\xi_{\mu}\xi^{\mu}=(U+X_{2})^{2}+a^{2}(V^{2}-X_{5}^{2})
+b_{1}^{2}(X_{6}^{2}+X_{7}^{2})
+\cdots+ b_{\frac{n-4}{2}}^{2}(X_{n}^{2}+X_{n+1}^{2}),
\end{equation}
for type $III_{0(b)}$ we have
\begin{equation}
\xi=-J_{01}+J_{02}+b_{1}J_{45}+b_{2}J_{67}+\cdots
+b_{\frac{n-2}{2}}J_{nn+1},
\end{equation}
with norm
\begin{equation}
\xi_{\mu}\xi^{\mu}=-(V+X_{2})^{2}+b_{1}^{2}(X_{4}^{2}+X_{5}^{2})
+\cdots+b_{\frac{n-2}{2}}^{2}(X_{n}^{2}+X_{n+1}^{2}),
\end{equation}
and for type $III_{0(c)}$ we have
\begin{equation}
\xi=-J_{13}+J_{23}+b_{1}J_{45}+b_{2}J_{67}+\cdots+b_{\frac{n-2}{2}}J_{nn+1},
\end{equation}
with norm
\begin{equation}
\xi_{\mu}\xi^{\mu}=(V+X_{2})^{2}+b_{1}^{2}(X_{4}^{2}+X_{5}^{2})
+b_{2}^{2}(X_{6}^{2}+X_{7}^{2})+\cdots+b_{\frac{n-2}{2}}^{2}(X_{n}^{2}+X_{n+1}^{2}).
\end{equation}
This last case is spacelike everywhere away from the subspace $V +
X_2=0$, $X_i = 0, i \geq 4$, where it is null. Note that
$III_{0(c)}$ includes $III_{0(a)}$ with $a=0$ as a special case. 
Finally, for type $V$ we have
\begin{equation}
\xi=-J_{01}-J_{02}-J_{13}-J_{14}-J_{23}+J_{24}+
b_{1}J_{67}+\cdots+b_{\frac{n-4}{2}}J_{nn+1},
\end{equation}
with norm
\begin{eqnarray}
\xi_{\mu}\xi^{\mu}&=&(V+X_{2})^{2}-2X_{4}(U+X_{3})
\nonumber \\ &&+b_{1}^{2}(X_{6}^{2}+
X_{7}^{2})+\cdots+b_{\frac{n-4}{2}}^{2}(X_{n}^{2}+X_{n+1}^{2}).
\end{eqnarray}

When $n$ is odd, the possibilities are slightly different:
\begin{itemize}
\item
Type $I$
\eqnarray
&\mathbb{C}& \qquad \lambda_{c}~(--++)\qquad+\lambda_{0}~(+)\qquad+\frac{n-3}{2}\lambda_{i}
~(+_{1}+_{2}\cdots+_{n-3}), \nonumber\\
&\mathbb{R}& \qquad 2\lambda_{r}~(--++)\qquad+\lambda_{0}~(+)\qquad+\frac{n-3}{2}\lambda_{i}
~(+_{1}+_{2}\cdots+_{n-3}), \nonumber\\
&\mathbb{I}& \qquad \frac{n+1}{2}\lambda_{i}
~(--+_{1}+_{2}\cdots+_{n-1})\qquad+\lambda_{0}~(+), \nonumber \\
&\mathbb{R}(0)& \qquad \lambda_{r}~(-+)\qquad+\lambda_{0}~(-)\qquad+\frac{n-1}{2}\lambda_{i}
~(+_{1}+_{2}\cdots+_{n-1}) \nonumber.
  \endeqnarray

\item
 Type $II$
\eqnarray
&\mathbb{R}& \qquad \lambda^{II}_{r}~(--++)\qquad+\lambda_{0}~(+)\qquad+\frac{n-3}{2}\lambda_{i}
~(+_{1}+_{2}\cdots+_{n-3}), \nonumber\\
&\mathbb{I}& \qquad \lambda^{II}_{i}~(--++)\qquad+\lambda_{0}~(+)\qquad+ \frac{n-3}{2}\lambda_{i}
~(+_{1}+_{2}\cdots+_{n-3}) \nonumber.
\endeqnarray
\item
Type $III$
\eqnarray
&\mathbb{I}& \qquad\lambda^{III}_{i}~(--++++)\qquad+\lambda_{0}~(+)\qquad+
 \frac{n-5}{2}\lambda_{i}
~(+_{1}+_{2}\cdots+_{n-5}), \nonumber\\
&0&(a)\qquad\lambda^{III}_{0}~(-++)\qquad+\lambda_{r}~(-+)
\qquad+ \frac{n-3}{2}\lambda_{i}
~(+_{1}+_{2}\cdots+_{n-3}), \nonumber\\
&0&(b)\qquad\lambda^{III}_{0}~(--+)\qquad+ \frac{n-1}{2}\lambda_{i}
~(+_{1}+_{2}\cdots+_{n-1}), \nonumber
\endeqnarray
\item
Type $V$
\eqnarray
\qquad\lambda^{V}_{0}~(--+++)\qquad+ \frac{n-3}{2}\lambda_{i}
~(+_{1}+_{2}\cdots+_{n-3}) \nonumber.
\endeqnarray
\end{itemize}

For the cases which occur for both even and odd $n$, the difference
between the two cases is just that for either even or odd $n$, there
is a direction which does not participate in the quotient; that is,
they differ by a factor of $\lambda_0$. It is therefore not worth
repeating the expressions for the Killing vectors in these cases for
$n$ odd. For the one new case, type $I_{\mathbb{R}(0)}$, the Killing
vector is
\begin{equation}
\xi=aJ_{12}+b_{1}J_{34}+b_{2}J_{56}+\cdots
+b_{\frac{n-1}{2}}J_{nn+1},
\end{equation}
with norm
\begin{equation}
\xi_{\mu}\xi^{\mu}=a^{2}(V^{2}-X_{2}^{2})+b_{1}^{2}(X_{3}^{2}+X_{4}^{2})+
b_{2}^{2}(X_{5}^{2}+X_{6}^{2})+\cdots+
b_{\frac{n-1}{2}}^{2}(X_{n}^{2}+X_{n+1}^{2}).
\end{equation}
For $a=0$, this is the same as type $I_{\mathbb{I}}$ with $b_1=0$ in
odd dimension. It is spacelike away from $X_i = 0, i \geq 3$, which is
an axis where the Killing vector degenerates. 

This completes the basic classification of different one-parameter
subgroups of $SO(2,n)$, which is the central result of our
paper. Most of the quotients determined by these Killing vectors will
have causal pathologies, so they are not of great physical
interest. The identification and description of the physically
interesting cases is the subject of a companion
paper~\cite{collab}. 

To conclude this paper, we briefly describe how convenient coordinate
systems can be defined on AdS$_{n+1}$ based on the construction of the
quotients out of invariant subspaces. These coordinate systems are
quite useful in understanding the relation between quotients for
different values of $n$ and in working out their physics. 

We have observed that the Killing vector describing each distinct type
of quotient naturally decomposes into an $SO(2,k)$ Killing vector,
with $k \leq 4$, and a series of $SO(2)$ rotations in independent
planes. This decomposition can be made explicit if we work in a
suitable coordinate system. For most types, $\xi$ can be decomposed in
terms of an $SO(2,2)$ Killing vector acting on the coordinates
$U,V,X_2,X_3$ and rotations acting on the remaining $X_i$ coordinates,
$i=4,\ldots,n+1$. We can then construct a suitable coordinate system
on AdS$_{n+1}$ (for $n \geq 3$) by writing
\begin{equation}
U = \cosh \chi \cosh \rho \cos t, \quad V = \cosh \chi \cosh \rho \sin
t,
\end{equation}
\begin{equation}
X_2 = \cosh \chi \sinh \rho \cos \phi, \quad X_3 = \cosh \chi \sinh
\rho \sin \phi,
\end{equation}
\begin{equation}
X_i = \sinh \chi x_i, 
\end{equation}
where $i = 4,\ldots,n+1$, and $x_i^2 =1$, so the metric on AdS$_{n+1}$
is 
\begin{eqnarray} \label{adsfol}
ds^2 &=& \cosh^2 \chi (-\cosh^2 \rho dt^2 + d\rho^2 + \sinh^2 \rho
d\phi^2) + d\chi^2 + \sinh^2 \chi d\Omega_{n-3} \\ &=& \cosh^2 \chi
ds^2_{AdS_3} + d\chi^2 + \sinh^2 \chi d\Omega_{n-3}. \nonumber
\end{eqnarray}
In this coordinate system, we can write $\xi = \xi_3 + \xi_r$, where
$\xi_3$ acts only on the AdS$_3$ part, while the $\xi_r$ is a rotation
acting on the unit sphere $S^{n-3}$. Furthermore, $\xi_3$ is precisely
the Killing vector associated to the same type of quotient in the
analysis of~\cite{btz2}. Similar coordinate systems can be introduced
in the remaining two cases, writing AdS$_{n+1}$ in terms of an AdS$_4
\times S^{n-4}$ or AdS$_5 \times S^{n-5}$ foliation. We will not
repeat the details of the coordinate transformation, which are quite
similar to the above case. 

The coordinate system (\ref{adsfol}) also gives us an interesting
description of the asymptotic boundary; taking the limit $\chi \to
\infty$ and conformally rescaling by a factor of $e^{-2\chi}$, we
can describe the asymptotic boundary in terms of AdS$_3 \times
S^{n-3}$ coordinates;
\begin{equation}
ds^2_{\Sigma} = (-\cosh^2 \rho dt^2 + d\rho^2 + \sinh^2 \rho
d\phi^2) + d\Omega_{n-3}.
\end{equation}
This description is related to the usual Einstein Static Universe
(ESU) metric $\mathbb{R} \times S^{n-1}$ on the conformal boundary of
AdS$_{n+1}$,
\begin{equation}
\tilde{ds}^2_{\Sigma} = -dt^2 + d\theta^2 + \sin^2 \theta d\phi^2 +
\cos^2 \theta d\Omega_{n-3},
\end{equation}
by a coordinate transformation $\cosh \rho = 1/\cos \theta$ and a
conformal rescaling $ds^2_{\Sigma} = \cosh^2 \rho
\tilde{ds}^2_{\Sigma}$. Hence, the AdS$_3 \times S^{n-3}$
coordinates cover all of the $S^{n-1}$ in the ESU except for one pole. 

These coordinatizations show that the action of a given quotient on
AdS$_{n+1}$ can be simply expressed in terms of the action of the
corresponding quotient on AdS$_3$ (or AdS$_4$ or AdS$_5$) subspaces of
the AdS$_{n+1}$ together with rotations in an orthogonal sphere. In
addition, the action of the quotient on the boundary of AdS$_{n+1}$ for
$n > 2$ ($n>3$, $n>4$ respectively) is also expressed in terms of the
action on the bulk of the lower-dimensional space. This observation
will be used extensively in the study of the physics of these
quotients in~\cite{collab}.  

The main purpose of this paper was to explore the extension of the
classification of one-parameter quotients of AdS$_d$, discussed
in~\cite{btz2,hp} for the cases $d=3,4$, to the general case. This
extension proved to be reasonably direct. Perhaps surprisingly, there
was little novelty in the general analysis; almost all the cases that
appear for general $d$ have appeared already for $d=3$~\cite{btz2} or
$4$~\cite{hp}. The one exception, type $III_{\mathbb{I}}$, extends a
particular quotient considered for the case $d=5$ in~\cite{simon}.

\medskip
\centerline{\bf Acknowledgements}
\medskip

  We thank Jose Figueroa-O'Farrill and Joan Simon for making the
  results of their paper \cite{fofs2} known to us before publication.
  OM is supported by the PPARC. SFR is supported by the EPSRC.

\providecommand{\href}[2]{#2}\begingroup\raggedright\endgroup


\begin{thebibliography}{10}

\bibitem{fofs}
J.~Figueroa-O'Farrill and J.~Simon, ``Generalized supersymmetric fluxbranes,''
  JHEP {\bf 12} (2001) 011,
\href{http://xxx.lanl.gov/abs/hep-th/0110170}{{\tt hep-th/0110170}}.

\bibitem{lms1}
H.~Liu, G.~Moore, and N.~Seiberg, ``Strings in a time-dependent orbifold,''
  JHEP {\bf 06} (2002) 045,
\href{http://xxx.lanl.gov/abs/hep-th/0204168}{{\tt hep-th/0204168}}.

\bibitem{lms2}
H.~Liu, G.~Moore, and N.~Seiberg, ``Strings in time-dependent orbifolds,'' JHEP
  {\bf 10} (2002) 031,
\href{http://xxx.lanl.gov/abs/hep-th/0206182}{{\tt hep-th/0206182}}.

\bibitem{fabmcg}
M.~Fabinger and J.~McGreevy, ``On smooth time-dependent orbifolds and null
  singularities,'' JHEP {\bf 06} (2003) 042,
\href{http://xxx.lanl.gov/abs/hep-th/0206196}{{\tt hep-th/0206196}}.

\bibitem{lawrence}
A.~Lawrence, ``On the instability of {3D} null singularities,'' JHEP {\bf 11}
  (2002) 019,
\href{http://xxx.lanl.gov/abs/hep-th/0205288}{{\tt hep-th/0205288}}.

\bibitem{horpol}
G.~T. Horowitz and J.~Polchinski, ``Instability of spacelike and null orbifold
  singularities,'' Phys. Rev. D {\bf 66} (2002) 103512,
\href{http://xxx.lanl.gov/abs/hep-th/0206228}{{\tt hep-th/0206228}}.

\bibitem{mald}
J.~M. Maldacena, ``The large {N} limit of superconformal field theories and
  supergravity,'' Adv. Theor. Math. Phys. {\bf 2} (1998) 231--252,
\href{http://xxx.lanl.gov/abs/hep-th/9711200}{{\tt hep-th/9711200}}.

\bibitem{magoo}
O.~Aharony, S.~S. Gubser, J.~M. Maldacena, H.~Ooguri, and Y.~Oz, ``Large {N}
  field theories, string theory and gravity,'' Phys. Rept. {\bf 323} (2000)
  183--386,
\href{http://xxx.lanl.gov/abs/hep-th/9905111}{{\tt hep-th/9905111}}.

\bibitem{ban1}
M.~Banados, ``Constant curvature black holes,'' Phys. Rev. D {\bf 57} (1998)
  1068--1072,
\href{http://xxx.lanl.gov/abs/gr-qc/9703040}{{\tt gr-qc/9703040}}.

\bibitem{ban2}
M.~Banados, A.~Gomberoff, and C.~Martinez, ``Anti-de {S}itter space and black
  holes,'' Class. Quant. Grav. {\bf 15} (1998) 3575--3598,
\href{http://xxx.lanl.gov/abs/hep-th/9805087}{{\tt hep-th/9805087}}.

\bibitem{simon}
J.~Simon, ``Null orbifolds in {AdS}, time dependence and holography,'' JHEP
  {\bf 10} (2002) 036,
\href{http://xxx.lanl.gov/abs/hep-th/0208165}{{\tt hep-th/0208165}}.

\bibitem{btz2}
M.~Banados, M.~Henneaux, C.~Teitelboim, and J.~Zanelli, ``Geometry of the (2+1)
  black hole,'' Phys. Rev. D {\bf 48} (1993) 1506--1525,
\href{http://xxx.lanl.gov/abs/gr-qc/9302012}{{\tt gr-qc/9302012}}.

\bibitem{hp}
S.~Holst and P.~Peldan, ``Black holes and causal structure in anti-de {S}itter
  isometric spacetimes,'' Class. Quant. Grav. {\bf 14} (1997) 3433--3452,
\href{http://xxx.lanl.gov/abs/gr-qc/9705067}{{\tt gr-qc/9705067}}.

\bibitem{fofs2}
J.~Figueroa-O'Farrill and J.~Simon, ``Supersymmetric {Kaluza-Klein reductions
  of AdS} backgrounds,''
 \href{http://xxx.lanl.gov/abs/hep-th/0401206}{{\tt hep-th/0401206}}.

\bibitem{collab}
J.~Figueroa-O'Farrill, O.~Madden, S.~F. Ross, and J.~Simon, ``Quotients of
  {AdS$_{p+1} \times S^q$}: causally well-behaved spaces and black
  holes,'', \href{http://xxx.lanl.gov/abs/hep-th/0402094}{{\tt hep-th/0402094}}.

\bibitem{btz1}
M.~Banados, C.~Teitelboim, and J.~Zanelli, ``The black hole in
  three-dimensional space-time,'' Phys. Rev. Lett. {\bf 69} (1992) 1849--1851,
\href{http://xxx.lanl.gov/abs/hep-th/9204099}{{\tt hep-th/9204099}}.

\end{thebibliography}
\end{document}